\documentclass[usenatbib]{mnras}
\usepackage{epsf,graphics,graphicx}
\usepackage{amsmath}
\usepackage{amssymb,latexsym,mathrsfs, bm}
\usepackage{color}
\usepackage{float}
\usepackage{natbib}
\usepackage{times}
\usepackage{pdflscape}
\usepackage{longtable}

\newcommand{\HI}{HI}

\title[HI abundances and clustering]{A halo model for cosmological neutral hydrogen : abundances and clustering}

\author[Padmanabhan, Refregier and Amara]{Hamsa
Padmanabhan\thanks{Electronic address: hamsa.padmanabhan@phys.ethz.ch},
Alexandre Refregier\thanks{Electronic address:
{alexandre.refregier@phys.ethz.ch}}, and
Adam Amara\thanks{Electronic address: {adam.amara@phys.ethz.ch}
}\\
Institute for Astronomy, ETH Zurich, Wolfgang-Pauli-Strasse 27, CH-8093 Z\"{u}rich, Switzerland}

\begin{document}
\date{ }
\maketitle

\begin{abstract}

We extend the results of previous analyses towards constraining the abundance and clustering of post-reionization ($z \sim 0-5$) neutral hydrogen (HI) systems using a halo model framework. We work with a comprehensive HI dataset including the small-scale clustering, column density and mass function of HI galaxies at low redshifts, intensity mapping measurements at intermediate redshifts and the UV/optical observations  of Damped Lyman Alpha (DLA) systems at higher redshifts. We use a Markov Chain Monte Carlo (MCMC) approach to constrain the parameters of the best-fitting models, both for the HI-halo mass relation and the HI radial density profile. We find that a radial exponential profile results in a good fit to the low-redshift HI observations, including the clustering and the column density distribution.  The form of the profile is also found to match the high-redshift  DLA observations, when used in combination with a three-parameter HI-halo mass relation and a redshift evolution in the HI concentration.  The halo model predictions are in good agreement with the observed HI surface density profiles of low-redshift galaxies, and the general trends in the the impact parameter and covering fraction observations of high-redshift DLAs. We provide convenient tables summarizing the best-fit halo model predictions.
\end{abstract}


\section{Introduction}
Mapping the intensity fluctuations of neutral hydrogen (HI) in the post-reionization phase of the universe (redshifts 0-5) promises stringent constraints on cosmology, large-scale structure and the evolution of the intergalactic medium. 
HI gas in galaxies acts as a tracer of the underlying dark matter in the absence of complicated reionization astrophysics, and is hence a valuable tool to study nonlinear effects in the matter power spectrum. The three-dimensional information contained in the redshifted HI 21-cm line potentially allows for probing much larger comoving volumes than optical galaxy surveys, and may therefore improve the precision in the measurement of the cosmological parameters \citep[e.g.,][]{bull2014} beyond that achieved by currently available observations. This can also be used place constraints on models of dark energy and modified gravity \citep[e.g.,][]{chang10, hall2013}.

In 21-cm intensity mapping, one aims to map out the distribution of HI without  resolving individual galaxies \citep[e.g.][]{santos2015}. This approach allows for a statistical study of the intensity fluctuations of HI and their evolution across cosmic time, through the 21-cm power spectrum $P_{\rm HI} (k,z)$. The two important ingredients in the power spectrum of intensity fluctuations are the neutral hydrogen density parameter, $\Omega_{\rm HI} (z)$ and the  bias parameter of HI relative to dark matter, $b_{\rm HI}(k,z)$, both of which are expected to evolve across redshifts in the post-reionization universe. The above quantities can be estimated if the underlying HI-halo mass relation (HIHM) is known, which provides an estimate of the average mass of HI, $M_{\rm HI} (M,z)$ contained in a dark matter halo of mass $M$ at redshift $z$. To quantify the small-scale and clustering behaviour, one also needs knowledge of the radial density profile of HI, $\rho_{\rm HI} (r)$ as a function of the distance $r$ from the halo. 

Existing measurements of the HI-based observables  include the 21-cm emission line data at low redshifts \citep[$z \sim 0$; ][]{zwaan05, zwaan2005a, martin10, martin12, braun2012}, the intensity mapping constraints at moderate redshifts \citep[$z \sim 1$;][]{switzer13, wolz2015}, and the high redshift UV/optical observations of HI in Damped Lyman Alpha systems \citep[DLAs; $z \sim 2-5:$][]{rao06,prochaska09, noterdaeme12,fontribera2012, zafar2013}. Using the combined set of data, it is possible to place constraints on the form of the HIHM and the HI density profile, and their redshift evolution. Both analytical techniques \citep[e.g.,][]{marin2010,bagla2010, barnes2014, hptrcar2015}, and hydrodynamical simulations \citep[e.g.,][]{dave2013,rahmati2013, bird2014} have been used to constrain the evolution of the HIHM, from the combined set (or subsets) of these observations at different redshifts.

In \citet[][hereafter Paper I]{hpar2016},  we introduced a halo model framework that describes the distribution and evolution of \HI{} across redshifts, focusing on the large-scale observables such as the neutral hydrogen density $\Omega_{\rm HI} (z)$ and bias parameter $b_{\rm HI} (z)$, in analogy with the corresponding halo model formulations for dark matter and galaxy evolution. The halo model also described the evolution of the statistical properties of DLAs at high redshifts (the DLA column density distribution $f_{\rm HI}$, incidence $dN/dX$, clustering bias $b_{\rm DLA}$, and their contribution to the density parameter of neutral hydrogen, $\Omega_{\rm DLA}$). 

In the present work, we build upon the analysis of Paper I to also include small-scale clustering, quantified by the scale-dependent correlation function of HI-selected galaxies at low redshifts \citep{martin12}. We also include recent data from the column density distribution of DLAs at $z \sim 5$ \citep{crighton2015}. The scale dependence of the clustering of HI-selected galaxies has been measured from the results of the ALFALFA survey \citep{martin12, papastergis2013} at $z \sim 0$. This therefore places constraints on both the HIHM relation as well as the HI profile at low redshifts. We constrain the parameters of the HIHM relation and the HI profile using a Markov Chain Monte Carlo (MCMC) analysis with the \textsc{cosmohammer} package \citep{akeret2013}. We describe the best-fitting relations so derived and their implications for HI intensity mapping and galaxy evolution.

This paper is organized as follows. In the next section (Sec. \ref{sec:halomod}), we briefly summarize the main ingredients in the HI halo model used, and the parameters therein. In Sec. \ref{sec:formalism}, we review the formalism for   estimating the abundances and clustering of HI systems across $z \sim 0 - 5$ with the halo model, particularly the correlation function of HI systems from the 1-  and 2-halo terms of the HI power spectrum. We then provide a brief summary of the data used to constrain the halo model from the results of 21-cm emission line measurements, intensity mapping surveys and Damped Lyman Alpha (DLA) observations in Sec. \ref{sec:data}. We describe the best-fit parameters of the halo model obtained by fitting the data with a Bayesian Markov Chain Monte Carlo (MCMC) method, and the comparison to the data in Sec. \ref{sec:compare}. We summarize our results and discuss future prospects in a brief concluding section (Sec. \ref{sec:summary}).

 Throughout the paper, we assume a flat $\Lambda$CDM cosmology consistent with previous work: $\Omega_m = 0.281$, $\Omega_\Lambda = 0.719$, $h = 0.71$, $\Omega_b = 0.0462$, $\sigma_8 = 0.8$, $n_s = 0.96$ and $Y_p = 0.24$.

\section{The HI halo model}
\label{sec:halomod}
In this section, we provide a brief summary of a halo model for cosmological HI, which addresses small-scale clustering and extends the results of previous work (Paper I).

In analogy with the dark matter framework, we describe the abundance and clustering of cosmic HI with a HI-halo mass relation [$M_{\rm HI}(M)$] together with a profile function $\rho_{\rm HI} (r)$ which represents the radial distribution of HI within a halo. These are briefly outlined in the following subsections.

\subsection{The HI-halo mass relation}
The HI-halo mass relation (HIHM) quantifies how the HI mass and halo mass are related to each other and is important from the point of view of determining the typical host halo masses of HI galaxies. 
A number of functional forms have been adopted in the literature, here we use a simple, three - parameter  HIHM, along the lines of Paper I. 

The model can be described by an $M_{\rm HI} - M$ relation of the form:
\begin{eqnarray}
M_{\rm HI} (M) &=& \alpha f_{H,c} M \left(\frac{M}{10^{11} h^{-1} M_{\odot}}\right)^{\beta} \exp\left[-\left(\frac{v_{c0}}{v_c(M)}\right)^3\right] \nonumber \\
\end{eqnarray}
The above relation involves the three free parameters $\alpha$, $\beta$ and $v_{c,0}$:
(i) $\alpha$ is an overall normalization and represents the fraction of HI, relative to cosmic ($f_{\rm H,c}$) associated with a dark matter halo of mass $M$. (ii) $\beta$ is the logarithmic slope of the $M_{\rm HI} - M$ relation, and was set to unity in some of the previous analyses \citep{hptrcar2016, barnes2014}. However, the best-fitting value of $\beta$ was found to be less than unity in order to fit the observations of the HI mass function at $z \sim 0$ (Paper I). (iii) The cutoff $v_{c,0}$ represents the minimum virial velocity of a host halo able to host HI. {The results of simulations \citep[e.g.][ see also \citet{bagla2010}]{pontzen2008},
disfavour the assignment of HI gas to halos with virial velocities smaller than 30 km/s. This is attributed to the UV field which prohibits the efficient cooling of gas in lower mass haloes, as also argued in previous literature, \citet{rees1986, efstathiou1992, quinn1996}. The lower cutoff in the virial velocity is thus also a constraint on the efficiency of stellar feedback in shallow potential wells (as also discussed in \citet{barnes2014}). In previous work, the value of $v_{c,0}$ was set to 30 km/s at low redshifts \citep{bagla2010} and increased to $\sim 35-50$ km/s to fit the DLA data at higher redshifts \citep{barnes2014}}.

In addition to the above parameters, Paper I also involved a quantity $v_{c,1}$ as a high-mass virial velocity cutoff. However, the best-fit value of $v_{c,1}$ was found to be very high in that analysis ($v_{c,1} \sim 10000$ km/s), and hence is neglected in the present study.

\subsection{The HI radial density profile}
This function describes the distribution of HI in a dark matter halo of mass $M$, as a function of radial distance $r$ from the centre of the halo, and is as such analogous to the corresponding [e.g. Navarro-Frenk-White \citep[NFW;][]{navarro1997}] dark matter halo profile. The form of the profile can be constrained by direct observations at different redshifts as well as from the measurements of small scale clustering and bias. In addition, indirect constraints on the geometry of the HI distribution around high-redshift absorption systems come from the observations of DLA impact parameters and covering fractions as a function of column density \citep{rao2011, krogager2012, rudie2012,peroux2013}.

In observational and simulation studies \citep[e.g.,][]{obreschkow2009, wang2014} it has been found that HI-rich galaxies may be described by exponential disk profiles at low redshifts. In the present analysis, we study the HI halo model with an exponential profile in the radial direction.\footnote{Strictly speaking, an exponential surface density profile arises from a radial profile of the form $\rho(r) = \rho_0 K_0(r/r_s)$ where $K_0$ is the modified Bessel function; however, we work with a radial exponential here for simplicity. This  form of the profile also does not involve a bulge component, which is expected to be sub-dominant in the context of HI observations.} The profile function is described by:
\begin{equation}
\rho(r,M) = \rho_0 \exp(-r/r_s)
\label{rhodefexp}
\end{equation}
In the above expressions,  $r_s$ is the scale radius of the dark matter halo, defined as $r_s \equiv R_v(M)/c_{\rm HI}(M,z)$, where $R_v(M)$ is the virial radius of the dark matter halo of mass $M$. The $c_{\rm HI}(M,z)$ is the concentration parameter of the HI, analogous to the corresponding concentration parameter of dark matter, and defined as \citep{maccio2007}:
\begin{equation}
 c_{\rm HI}(M,z) =  c_{\rm HI, 0} \left(\frac{M}{10^{11} M_{\odot}} \right)^{-0.109} \frac{4}{(1+z)^{\gamma}}.
\end{equation} 
The constant $\rho_0$ in Eq. (\ref{rhodefexp}) is fixed by normalizing the profile within the virial radius $R_v$ to be equal to $M_{\rm HI}$. Hence, the two free parameters in the HI density profile are $c_{\rm HI, 0}$ and $\gamma$.
 For the concentration parameters in the regime of interest ($c_{\rm HI} > 10$), this can be well approximated as:
\begin{equation}
\rho_0 = M_{\rm HI} (M) /(8 \pi r_s^3)
\end{equation}
We use this form of the profile for the analyses in the main text. In previous work \citep[][Paper I]{barnes2014, hptrcar2016}, an altered NFW profile was considered, which was found to be a good fit to multiphase gas in simulations \citep{maller2004}. We describe the results obtained with the modified NFW form of the profile in the Appendix.

\begin{figure*}
\begin{center}
\includegraphics[scale=0.6, width = \textwidth]{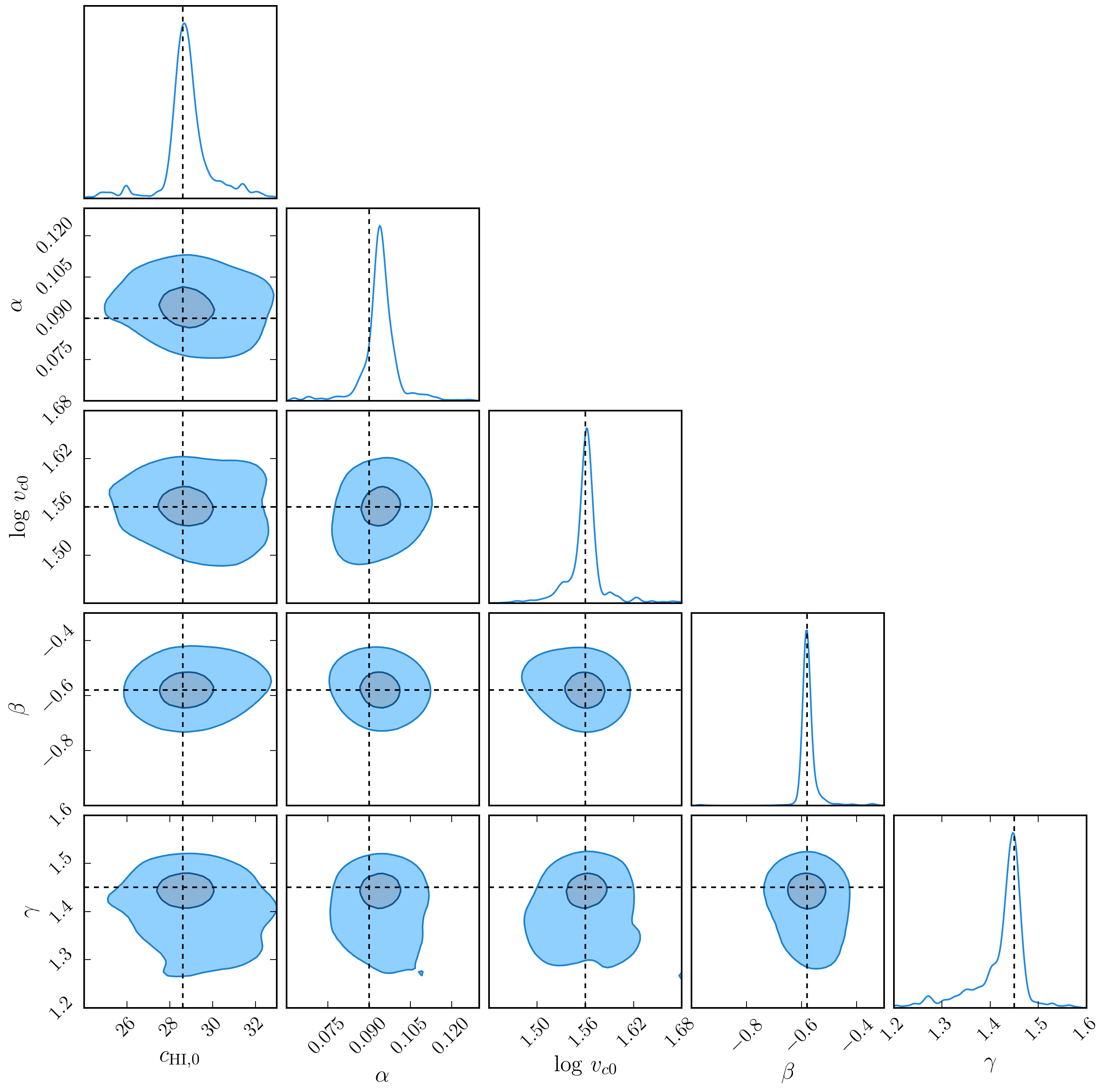}
\end{center}
\caption{Parameter space showing the constraints from the MCMC analysis. Contours indicate 1- and 2$\sigma$ confidence levels. The crosshairs indicate the maximum likelihood estimate of the parameters (the best-fitting value). The marginal distributions of each parameter are shown in the diagonal panels. }
\label{fig:cornerplot}
\end{figure*}

\section{Abundances and clustering}
\label{sec:formalism}
Given the two ingredients in the halo model (i.e. the radial distribution of HI, and the HI-halo mass relation), we can compute various quantities related to HI evolution: the neutral hydrogen density parameter, $\Omega_{\rm HI} (z)$, the HI bias $b_{\rm HI} (z)$ and the HI mass function  $\phi(M_{\rm HI}, z)$, as well as the quantities related to high-redshift DLAs: the column density distribution $f_{\rm HI} (z)$, the DLA incidence $dN/dX$, the DLA neutral hydrogen density parameter $\Omega_{\rm DLA} (z)$, and the large scale clustering bias of the DLAs, $b_{\rm DLA} (z)$. The detailed expressions for these quantities are specified in previous work \citep[Paper I,][]{hptrcar2015, hpgk2016}, and are summarized in Table \ref{table:model}. In the present work, we also consider the small-scale clustering of HI and its dependence on the free parameters.

\begin{table*}
\centering
\caption{Summary of the equations used to derive the various quantities in the HI halo model. The mass of the hydrogen atom is $m_H$, and other symbols have their usual meanings.}
\label{table:model}
\begin{tabular}{llll}
\hline 
Observable & Expression \\
\hline \\
   $N_{\rm HI}(s)$ & $(2/m_H) \int_0^{\sqrt{R_v(M)^2 - s^2}} \rho_{\rm HI} (r = \sqrt{s^2 + l^2}) \ dl$         \\
   \\
   \\
   $dN/dX$ &  $(c/H_0) \int_0^{\infty} n(M,z) \sigma_{\rm DLA}(M,z) \ dM$;  $\sigma_{\rm DLA}(M)$ = $\pi s_*^2$; $ \ N_{\rm HI} (s_*)=10^{20.3} {\rm cm}^{-2}$ \\
   \\  
   \\
   $b_{\rm DLA} (z)$ & $\int_{0}^{\infty} dM n (M,z) b(M,z) \sigma_{\rm DLA} (M,z)/\int_{0}^{\infty} dM n (M,z) \sigma_{\rm DLA} (M,z)$ \\
\\
\\
$f(N_{\rm HI}, z)$ & $ (c/H_0)\int_0^{\infty} n(M,z) \left|\frac{d \sigma}{d N_{\rm HI}} (M,z) \right| \ dM$  ; \ \  $d \sigma/d N_{\rm HI}$ =  $2 \pi \ s \ ds/d N_{\rm HI}$
\\
\\
$\Omega_{\rm DLA}(N_{\rm HI}, z)$   & $(m_H H_0/c \rho_{c,0}) \int_{10^{20.3}}^{\infty} f_{\rm HI}(N_{\rm HI}, z) N_{\rm HI} d N_{\rm HI}$
\\
\\
$\phi(M_{\rm HI}, z)$ & $n (M,z)  \left|\frac{d M}{d M_{\rm HI}} (M,z) \right|$ 
\\
\\
$\Omega_{\rm HI} (z)$ & $\frac{1}{\rho_{c,0}} \int_0^{\infty} n(M, z) M_{\rm HI} (M,z) dM$
\\
\\
$b_{\rm HI} (z)$ & $\int_{0}^{\infty} dM n(M,z) b (M,z) M_{\rm HI} (M,z)/\int_{0}^{\infty} dM n(M,z) M_{\rm HI} (M,z)$
\\
   \\
\hline \\    
\end{tabular}
\end{table*}

For quantifying clustering, we use the normalized Fourier transform of the \HI{} density profile, which is given by:
\begin{equation}
u_{\rm HI}(k|M) = \frac{4 \pi}{M_{\rm HI} (M)} \int_0^{R_v} \rho_{\rm HI}(r) \frac{\sin kr}{kr} r^2 \ dr
\end{equation}
where the normalization is to the total \HI{} mass in the halo, and the profile is assumed truncated at the virial radius of the host halo.
The expression for the normalized Fourier transform of the exponential HI profile is given by:
\begin{equation}
u_{\rm HI}(k|M) = \frac{4 \pi \rho_0 r_s^3 u_1(k|M)}{M_{\rm HI} (M)}
\end{equation}
where
\begin{equation}
u_1(k|M) = \frac{2}{(1 + k^2 r_s^2)^2}.
\end{equation}

The one- and two halo terms of the \HI{} power spectrum are then given by:
\begin{equation}
P_{\rm 1h, HI} =  \frac{1}{\bar{\rho}_{\rm HI}^2} \int dM \  n(M) \ M_{\rm HI}^2 \ |u_{\rm HI} (k|M)|^2
\end{equation}
and
\begin{equation}
P_{\rm 2h, HI} =  P_{\rm lin} (k) \left[\frac{1}{\bar{\rho}_{\rm HI}} \int dM \  n(M) \ M_{\rm HI} (M) \ b (M) \ |u_{\rm HI} (k|M)| \right]^2
\end{equation}
In both the above expressions, 
\begin{equation}
\bar{\rho}_{\rm HI} = \int dM n(M) M_{\rm HI} (M) 
\end{equation}
analogous to the corresponding quantities for dark matter. The dark matter mass function $n(M)$ is assumed to have the Sheth - Tormen form \citep{sheth2002}.

The \HI{} correlation function is then defined as:
\begin{equation}
\xi_{\rm HI} (r) = \frac{1}{2 \pi ^2} \int k^3 (P_{\rm 1h, HI} + P_{\rm 2h, HI}) \frac{\sin kr}{kr} \frac{dk}{k}
\end{equation}
We note that the above expression describes the mass-weighted correlation function of the \HI{} galaxies. The quantity that is usually measured in the \HI{} galaxy surveys  \citep[e.g. ALFALFA;][]{martin12}, is the number-weighted galaxy-galaxy correlation function, i.e.  $\xi_{\rm gg, HI} (r)$. However, it has been found that the number weighted correlation function in different mass bins of the ALFALFA data shows very little dependence on the galaxy HI mass \citep{papastergis2013}. Hence, in the comparison to data, we make the approximation that the number-weighted correlation function can be approximated by a combination of the mass-weighted clustering and a low-mass cutoff, along the lines of the current model. \footnote{We sub-sample the number-weighted correlation at regular intervals in order to mitigate the effects of the correlation between the individual data points.}

\section{Data}
\label{sec:data}
To constrain the free parameters of the model, we use the combination of the data from the low-redshift 21-cm observations and the higher redshift Damped Lyman Alpha (DLA) data, described in detail in \citet{hptrcar2015}. We add to this database the clustering data from the ALFALFA survey \citep{martin12} which gives $\xi_{\rm HI}(r)$ at $z \sim 0$,  the recent estimates of the column density distribution of Damped Lyman Alpha (DLA) absorbers at $z \sim 5$ \citep{crighton2015}, and the incidence of DLAs at $z \gtrsim 3$ \citep{zafar2013}. 

The resulting database of HI observables can thus be described by:

(i) $z \sim 0$: We consider the column density distribution of HI from the WHISP survey \citep{zwaan2005a}, the HI mass function from the HIPASS survey \citep{zwaan05} and the clustering of HI galaxies from the ALFALFA survey \citep{martin12}. We also use the large scale bias parameter of HI galaxies from ALFALFA \citep{martin12} and the DLA incidence $dN/dX$ measured from both the WHISP and the \citet{braun2012} surveys.

(ii) $z \sim 1$: We use the DLA incidence, $dN/dX$ and the column density distribution $f_{\rm HI}$ measured at $z \sim 0.6$ and 1.2 from the study of \citet{rao06}. We also consider the intensity mapping results, which constrain $\Omega_{\rm HI} b_{\rm HI}$ at $z \sim 0.8$ \citep{switzer13}. 

(iii) $z \sim 2.3$: We use the column density distribution $f_{\rm HI}$ at $z \sim 2.3$ \citep{noterdaeme12}, the neutral hydrogen density parameter $\Omega_{\rm DLA}$ \citep{zafar2013} and the clustering of DLAs systems, which constrains $b_{\rm DLA}$ at $z \sim 2.3$ \citep{fontribera2012}.

(iv) $z \geq 3$: We use the incidence $dN/dX$ of DLAs at $z \sim  3, 3.5$ and 4 \citep{zafar2013}, and the column density distribution of DLA systems measured at $z \sim 5$ \citep{crighton2015}.

The data considered are summarized in Table \ref{table:data}. The table also indicates the nature and details of the $1\sigma$ errors used, where available.
\begin{table}
\centering
\begin{tabular}{llll}
\hline
 Redshift & Data  & Reference  \\
\hline
0 & $f_{\rm HI}$ & WHISP; \citet{zwaan2005a}$^{*}$            \\
0     & $\phi_{\rm HI}$       &     HIPASS; \citet{zwaan05}$^{\dagger}$          \\
0    &  $\xi_{\rm HI}$, $b_{\rm HI}$     &   ALFALFA; \citet{martin12} $^{**}$      \\
0      & $dN/dX$  &   \citet{zwaan05}, \citet{braun2012}      \\
1 & $f_{\rm HI}$ & \citet{rao06}             \\
1     & $\Omega_{\rm HI} b_{\rm HI}$  & \citet{switzer13}       \\
1   &  $dN/dX$      &   \citet{rao06}       \\
2.3    & $f_{\rm HI}$ & \citet{noterdaeme12}$^{\dagger}$        \\
2.3 & $\Omega_{\rm DLA}$ & \citet{zafar2013} \\
2.3  & $b_{\rm DLA}$ & \citet{fontribera2012} \\
3 & $dN/dX$ & \citet{zafar2013}\\
3.5 & $dN/dX$ & \citet{zafar2013}\\
4 & $dN/dX$ & \citet{zafar2013}\\
5 & $f_{\rm HI}$ & \citet{crighton2015}$^{\dagger \dagger}$             \\
\hline \\   
\end{tabular}
\caption{Summary of the data used to constrain the HI halo model in this work.
$^{*}$ Errors indicate both, the uncertainties in the HI mass function normalisation \citep{zwaan2003}, and the counting statistics of the WHISP sample;
$^{\dagger}$ errors indicate the statistical Poisson component; 
 $^{**}$ errors indicate the on-diagonal terms from the full covariance matrix;
  $^{\dagger \dagger}$ errors indicate the statistical and systematic components. 
  }
  \label{table:data}
\end{table}

\section{MCMC analysis}
\label{sec:compare}

We can now constrain the free parameters of the model by using a Bayesian Markov Chain Monte Carlo (MCMC) method with the \textsc{cosmohammer} package \citep{akeret2013}. We approximate the likelihood to be in the form of a Gaussian:
\begin{equation}
\mathcal{L} =  \exp -\frac{1}{2}\sum \frac{(f_{\rm i, mod} - f_{\rm i,obs})^2}{\sigma_i^2} 
\end{equation}
where the $f_{\rm i, obs}$'s are the observed values of the data, the $f_{\rm i, mod}$'s are the model predictions, and the $\sigma_i$'s are the errors on the observed quantities, assumed to be independent.  

The parameters and their prior ranges are summarized in the first two columns of Table \ref{table:expbestfit}. We use flat, uniform priors for all 5 free parameters: three ($\alpha, v_{\rm c,0}$ and $\beta$) for the $M_{\rm HI} - M$ relation  and two ($c_{\rm HI}$ and $\gamma$) for the HI profile.  The prior ranges are chosen to be similar to those in Paper I.
We sample the likelihood using 20 random walkers for each for the 5 parameters and 200 iterations per random walker, making a total of 20000 iterations (after an initial burn-in phase of 30000 iterations).
 
The parameter space of the results is shown in Fig.  \ref{fig:cornerplot} using the \textsc{chainconsumer} routine \citep{Hinton2016}. The diagonal panels show the marginalized distributions for each of the parameters. The best-fitting values and the $1\sigma$ confidence intervals  are provided in the third column of Table \ref{table:expbestfit}.

\begin{table}
\centering
\caption{Priors and best-fitting values of the free parameters in the HI halo model.}
\label{table:expbestfit}
\begin{tabular}{llll}
\hline 
Parameter  & Prior & Best-fit \& Error (1$\sigma$) \\
\hline \\
$c_{\rm HI}$ &    [20, 400]   &   $28.65 \pm 1.76$              \\
$\alpha$     & [0.05, 0.5]      &     $0.09 \pm 0.01$             \\
log $v_{c,0}$    & [1.30, 1.90]      &   $1.56 \pm 0.04$          \\
$\beta$      &   [-1,3]    &     $-0.58 \pm 0.06$               \\
$\gamma$     &  [-0.9,2]     &       $1.45 \pm 0.04$     \\
\hline \\    
\end{tabular}
\end{table}

\subsection{Comparison to data}

{We present in this section a detailed comparison and discussions of the model predictions with the available data.}

\begin{figure}
\begin{center}
\includegraphics[scale=1, width = \columnwidth]{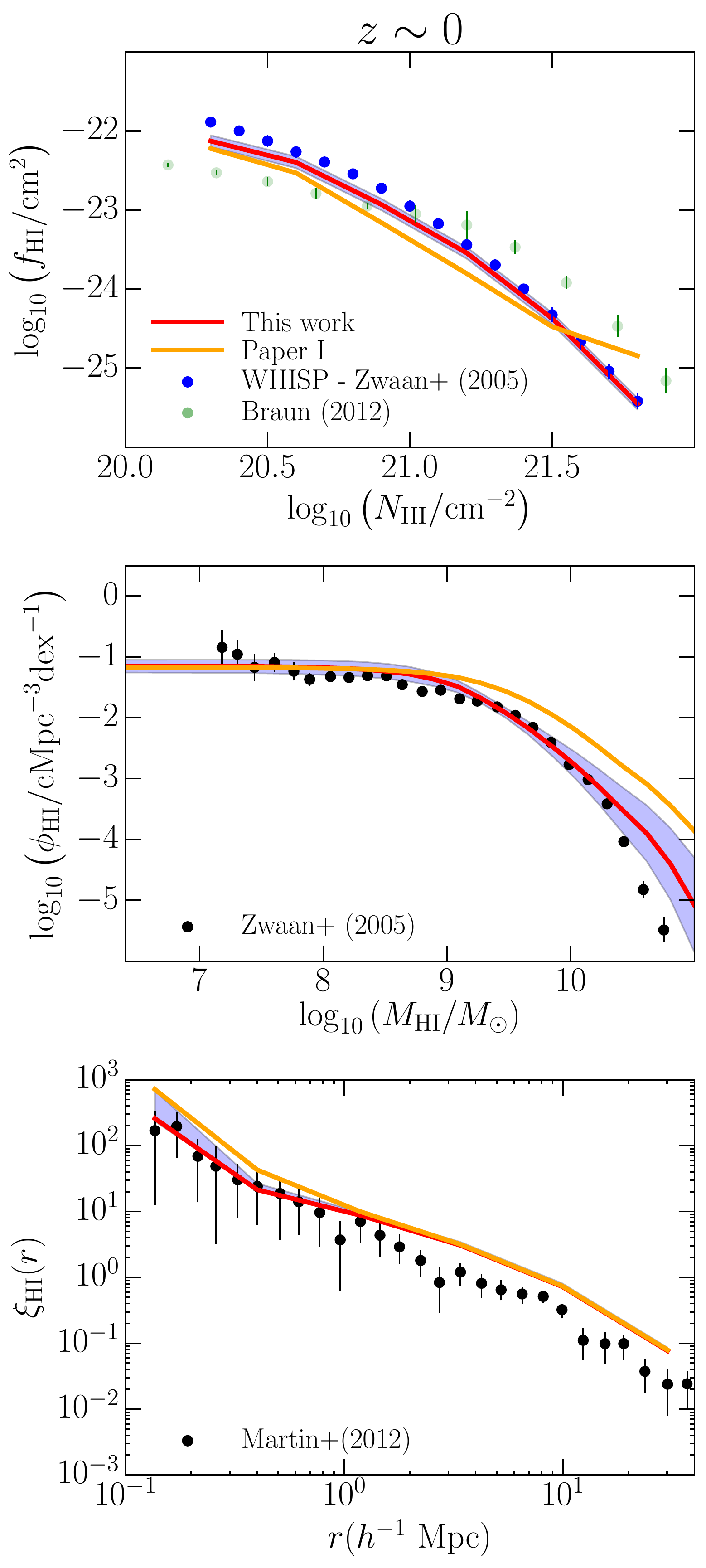}
\end{center}
\caption{Data and model predictions at $z \sim 0$. From top to bottom: (a) the column density distribution $f_{\rm HI}$ as measured from the WHISP \citep{zwaan2005a} and the \citet{braun2012} surveys, (b) the HI mass function $\phi_{\rm HI}$ from the results of the HIPASS \citep{zwaan05}  survey, and (c) the correlation function of HI-selected galaxies, $\xi_{\rm HI}(r)$ from the ALFALFA survey \citep{martin12}. The results from using the best-fit parameters in Paper I are shown in orange on each panel.}
\label{fig:clustermfexp}
\end{figure}

\begin{enumerate}

\item In Figure \ref{fig:clustermfexp}, we plot the comparison of the predictions of our model to the available data at $z \sim 0$. The uncertainty in the model predictions is indicated by the blue band in each panel. The top panel shows both the WHISP \citep{zwaan2005a} and the \citet{braun2012} data for the column density distribution, although the WHISP data is actually fitted. {The \citet{zwaan2005a} data uses 355 high-quality interferometric maps of nearby galaxies obtained with the Westerbork Synthesis Radio Telescope (WSRT) to derive the HI column density distribution function at $z \sim 0$. The \citet{braun2012} data uses high-resolution, opacity corrected images of the HI distribution in M31, M33 and the Large Magellanic Cloud (LMC) to estimate the column density distribution. Differences between these two datasets could possibly originate from the differences in the median beam sizes of the WHISP and the \citet{braun2012} observations, as well as the contribution of opacity corrections to the column density.}

The middle panel shows the \citet{zwaan05} data from the HIPASS (HI Parkes All-Sky Survey), {which uses 4315 detections of 21-cm line emission from  nearby galaxies to derive the HI mass function in the local universe ($z \sim 0$). The lower panel shows the correlation function of approximately 10150 HI-selected galaxies from the 40\% Arecibo Legacy Fast ALFA (ALFALFA) survey \citep{martin12} at $z \sim 0$, which is a measure of the clustering properties of HI galaxies as a function of scale. In all the panels, the model predictions are shown in red with the shaded region representing the associated error.}  At low $M_{\rm HI}$ values, the model predictions tend to slightly underestimate the HI mass function, with the opposite effect seen at high masses. This may indicate a possible tension between the HI mass function and the column density distribution at $z \sim 0$, though the effect is smaller compared to the case with an altered NFW profile, as seen from the results of Paper I (see also Appendix \ref{sec:appendix}). This may also point to evidence for the exponential profile to be favoured at low redshifts to describe the abundance and clustering of HI galaxies \citep[also indicated in, e.g.,][]{obreschkow2009,wang2014}. {We elaborate on this further in Sec. \ref{sec:summary}.}

The DLA column density distribution for higher redshifts is shown in the panels of Fig. \ref{fig:columndensityexp} along with the model predictions. The observations fitted are shown in black, from the measurements of \citet{rao06, noterdaeme12} and \citet{crighton2015} at redshifts 1, 2.3 and 5 respectively. In the third panel, the results from \citet{noterdaeme12} are also plotted in a lighter shade, which are roughly consistent with the $z \sim 5$ measurements but indicate evolution around column densities of $\log N_{\rm HI} \sim 21.2$ \citep{crighton2015}. {The overall similarity in the figures indicates that the column density distribution of DLA systems evolves only weakly from low to high redshifts. This is consistent with a series of results from the Sloan Digital Sky Survey \citep[][]{prochaska2005, prochaska09, noterdaeme09, noterdaeme12} which find evidence only for weak evolution in the mass density of HI over $z \sim 0 - 3$. This is in contrast to the strong evolution of the star-formation rate density over these epochs \citep[e.g.,][]{madau2014}. These two effects may be reconciled by the continuous replenishment of galactic HI consumed in star-formation by gas accreted from the IGM \citep[e.g.,][]{prochaska2005, crighton2015}.}

\begin{figure*}
\begin{center}
\includegraphics[scale=1, width = \textwidth]{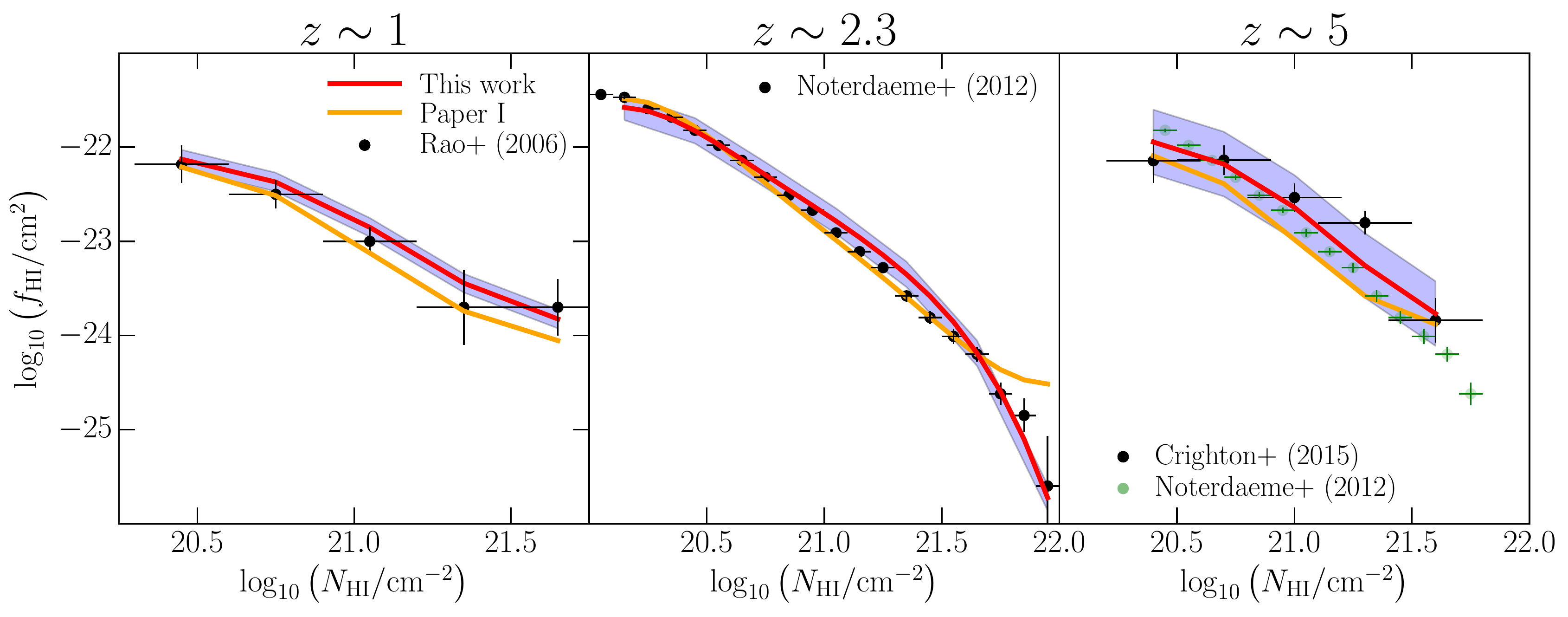}
\end{center}
\caption{The DLA column density distribution at redshifts $z > 0$, compared to the model predictions. From left to right: the column density distribution for DLAS at $z \sim 1$, $z \sim 2.3$, and $z \sim 5$ compared with the observations of \citet{rao06, noterdaeme12, crighton2015} respectively. The results from using the Paper I best-fit parameters are indicated in orange.}
\label{fig:columndensityexp}
\end{figure*}

Figure \ref{fig:omegabiasdndxexp} shows the comparison of the model predictions to the
{integrated DLA and 21-cm observables. The top panel shows the evolution of the neutral density parameter of DLAs,  $\Omega_{\rm DLA}$ over $z \sim 1-5$ from the results of \citet{rao06, zafar2013} and \citet{prochaska09}. The second panel shows the 21-cm intensity mapping measurement \citep{switzer13} of the product $\Omega_{\rm HI} b_{\rm HI}$ at $z \sim 1$ from combining the results of the auto-power spectrum from the Green Bank Telescope (GBT) with the 21cm - galaxy cross-correlation from the WiggleZ survey \citep{masui13}. The third panel shows the clustering, the HI galaxy bias $b_{\rm HI}$ at $z \sim 0$ from the ALFALFA survey, \citet{martin12} and the clustering bias of DLAs, $b_{\rm DLA}$ at $z \sim 2.3$ measured from the cross-correlation of the Lyman-alpha forest with DLAs in the BOSS survey \citep{fontribera2012}. The last panel shows the incidence of DLAs, $dN/dX$ over $z \sim 0-5$ from the results of \citet{zwaan2005a, braun2012, rao06} and \citet{zafar2013}.
 The model predictions are shown in red in each of the panels.} Again, all the observations are fairly well fit by the model predictions, though there is some difficulty in fitting the measured bias of DLAs at $z \sim 2.3$. 
\begin{figure}
\begin{center}
\includegraphics[scale=1, width = \columnwidth]{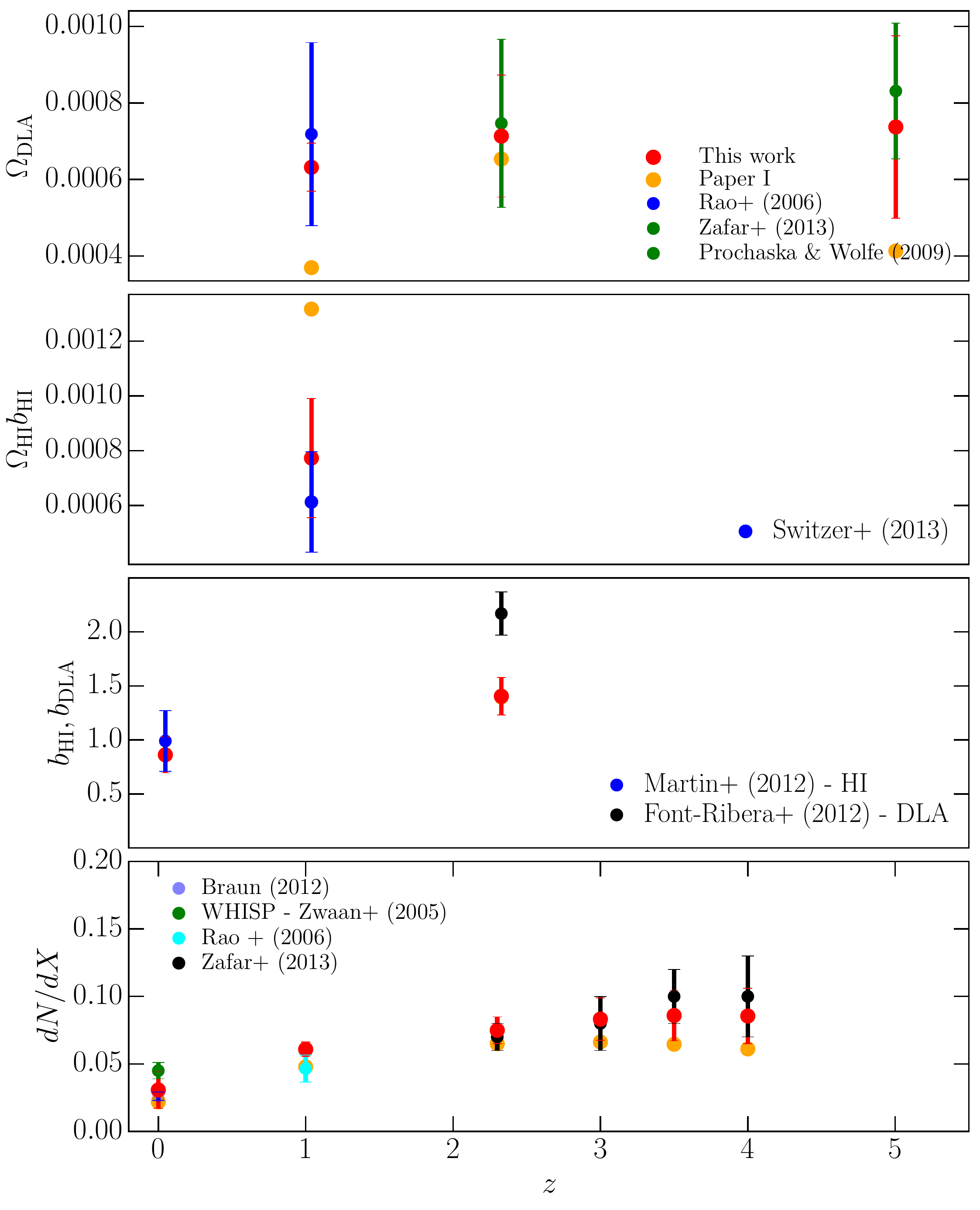}
\end{center}
\caption{From top to bottom: The observables $\Omega_{\rm DLA}$,  $\Omega_{\rm HI} b_{\rm HI}$ from intensity mapping experiments, the bias measurements $b_{\rm HI}$ and $b_{\rm DLA}$, and the DLA incidence $dN/dX$. Also shown are the model predictions (red) at the corresponding redshifts, and the results from Paper I which are indicated in orange.}
\label{fig:omegabiasdndxexp}
\end{figure}

In all the three Figures \ref{fig:clustermfexp}, \ref{fig:columndensityexp} and \ref{fig:omegabiasdndxexp}, the orange lines indicate the mean relation from the best-fit parameters of Paper I, showing that the results of Paper I are broadly consistent with the extended dataset including the clustering at low redshifts. Fig. \ref{fig:clustermfexp} indicates that the tension between the low-redshift column density distribution and the HI mass function can be reduced on using the exponential HI profile.

The Keck Baryonic Spectroscopic Survey  at redshifts 2-3 \citep[KBSS;][]{rudie2012} provides measurements of the covering fraction of DLAs located at one and two virial radii from their host dark matter haloes. These measurements may provide useful constraints on the geometry of the HI distribution in  high-density absorbers. We plot the predicted covering fraction of DLA systems at redshift 2.5 as a function of impact parameter $b$ from the DLA in Fig. \ref{fig:covfraction}, along with the observations. The covering fraction is computed for three representative host halo masses of $10^{11.5}$, $10^{12.1}$ and $10^{12.5} h^{-1} M_{\odot}$ respectively. Also plotted are the results from the cosmological zoom-in simulations with galactic winds from \citet[][pink crosses]{fumagalli2011} at the same redshift. We see that the model predictions are in good agreement with the results of the observations and simulations. The model predictions also favour an anti-correlation between the impact parameter and the column density distribution of high-redshift DLAs, as seen in several recent studies \citep[e.g.][]{peroux2013, krogager2012, rao2011}. 

\begin{figure}
\begin{center}
\includegraphics[scale=0.5, width = \columnwidth]{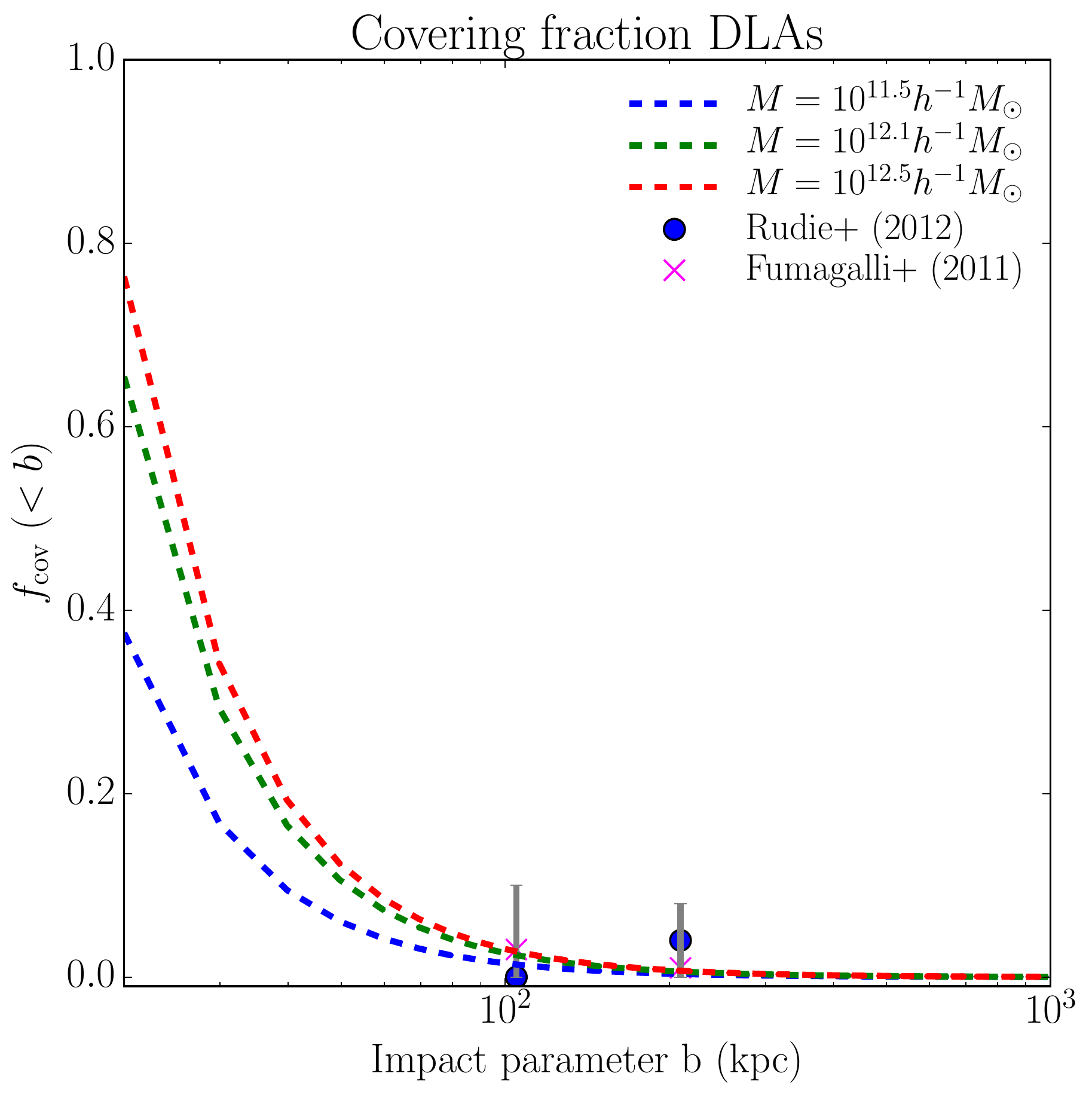}
\end{center}
\caption{The covering fraction as a function of impact parameter at $z \sim 2.5$, as predicted by the HI halo model, along with the observations from the Keck Baryonic Spectroscopic Survey \citep[KBSS;][blue circles]{rudie2012}  and the cosmological zoom-in simulations of \citet[][pink crosses]{fumagalli2011}. Results for three representative host halo masses are shown.}
\label{fig:covfraction}
\end{figure}

The model predictions at $z \sim 0$ can also be related to recent observational results which constrain the profile parameters  of gas-rich galaxies at low redshifts. In \citet{bigiel2012}, an analysis of 17 disk galaxies reveals a ``universal" exponential  surface density profile for low-redshift HI-rich disks. The present model predictions are found to be in good agreement with these findings. For host halo masses of $\sim 10^{11} M_{\odot}$, corresponding to the turnover scale in the HI-halo mass relation, the HI surface density at the characteristic optical radius is found to be about $6.9 \ M_{\odot} \mathrm{pc}^{-2}$, close to the typical values in the observations. This lends further support to the radial exponential profile parameters at low redshifts.

\end{enumerate}

\section{Summary and conclusions}
\label{sec:summary}
In this work, we have constructed a halo model formalism to describe the abundance and clustering of neutral hydrogen in the post-reionization universe ($z < 6$). In so doing, we have extended the results of previous analyses to incorporate small-scale clustering at $z \sim 0$, as well as higher redshift data beyond $z \sim 4$. We can summarize our main conclusions as follows:

\begin{enumerate}

\item The best-fitting model as predicted by Paper I is an overall good fit the extended dataset including the clustering at low redshifts. \footnote{This is also evidenced by the fact that the best-fitting parameters for the HI-halo mass relation with the modified NFW profile change very little even when the full dataset is taken into account, see Appendix \ref{sec:appendix}.}
\item The observational data at high- and low-redshifts favour an $M_{\rm HI} - M$ relation with an overall amplitude, slope and low-velocity cutoff, with the best-fitting parameters for these values summarized in Table \ref{table:final}. 
\item The combined set of low-redshift observations likely favours an exponential profile for the HI distribution, with evidence for evolution in the concentration parameter with increasing redshift, as summarized in Table \ref{table:final}. This form of the profile significantly reduces the tension (see Fig. \ref{fig:clustermfexp}) between the low-redshift column density distribution and the HI mass function observed (e.g., Paper I and \citet{hpgk2016}) with the modified NFW profile. Further, this form of the profile is also found to be a good fit to the high-redshift DLA data.
\item  {This is consistent with the picture of DLAs arising in large gaseous disks which were progenitors of present-day spiral galaxies \citep[e.g.,][]{wolfe1986}. Analysis of the kinematics and metal lines from DLAs \citep[e.g.,][]{prochaska1997} also supports the morphology of thick, rapidly rotating disks. \footnote{{However, there are arguments to show that the  velocity profiles could alternatively be consistent with DLAs arising as protogalactic clumps aggregating on haloes \citep[e.g.,][]{haehnelt1998}.}} Observations of local spiral galaxies indicate the radial surface brightness distribution to be well modelled by exponential disk profiles \citep[e.g.,][]{freeman1970}, and models of star formation are consistent with the assumption that cold gas does not change its distribution as it is converted into stars \citep[e.g.,][]{maller2000}. As also mentioned previously, the observed surface density profiles of HI galaxies at low redshifts are well fit by the exponential form \citep[e.g.,][]{wang2014, bigiel2012}.}

\item {We note that the low redshift HIHM relation is completely constrained by the form of the HI mass function, which strongly prefers a non-unity slope of the HIHM \citep{hpar2016, hpgk2016}. This leaves freedom only the form of the profile to jointly match the HIHM and the $z \sim 0$ column density distribution. An altered NFW profile, though a good match to high redshift DLA data \citep{hptrcar2016, barnes2014}, leads to some tension between the observed column density distribution and the HIHM at $z \sim 0$, which is reduced on using the exponential form of the profile.}

Fig. \ref{fig:massrelation} shows the derived $M_{\rm HI} - M$ relation based on the best-fit parameters. The  evolution of the profile $\rho(r)$ as a function of $z$ is shown in Fig. \ref{fig:profilerelation} for a fixed halo mass $M = 10^{12} h^{-1} M_{\odot}$.  
\begin{figure}
\begin{center}
\includegraphics[scale=0.5, width = \columnwidth]{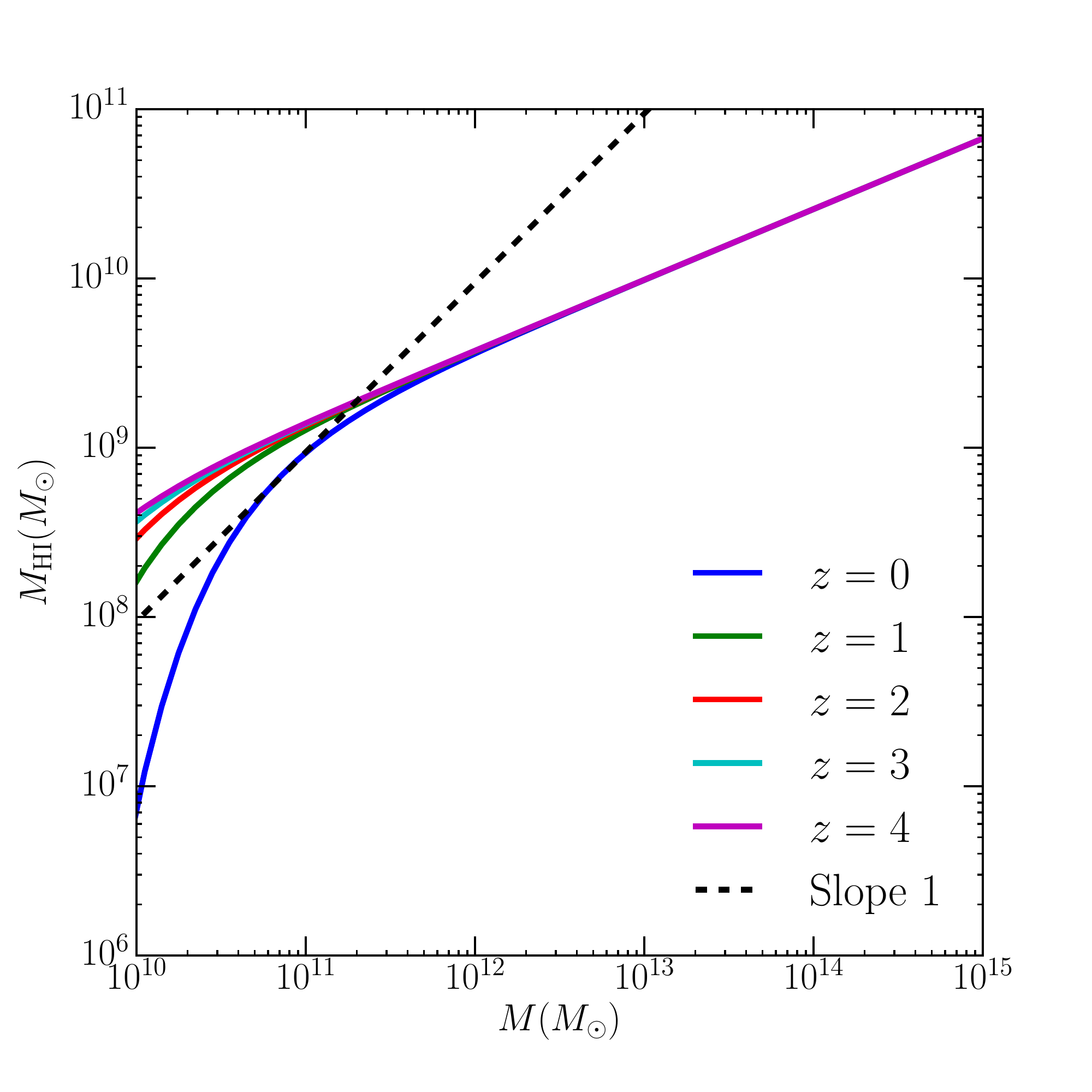}
\end{center}
\caption{The HI-halo mass relation across redshifts, with the best-fit halo model parameters. For comparison, the dashed line shows the relation with logarithmic slope unity, illustrating the effect of negative $\beta$.}
\label{fig:massrelation}
\end{figure}

\begin{figure}
\begin{center}
\includegraphics[scale=0.5, width = \columnwidth]{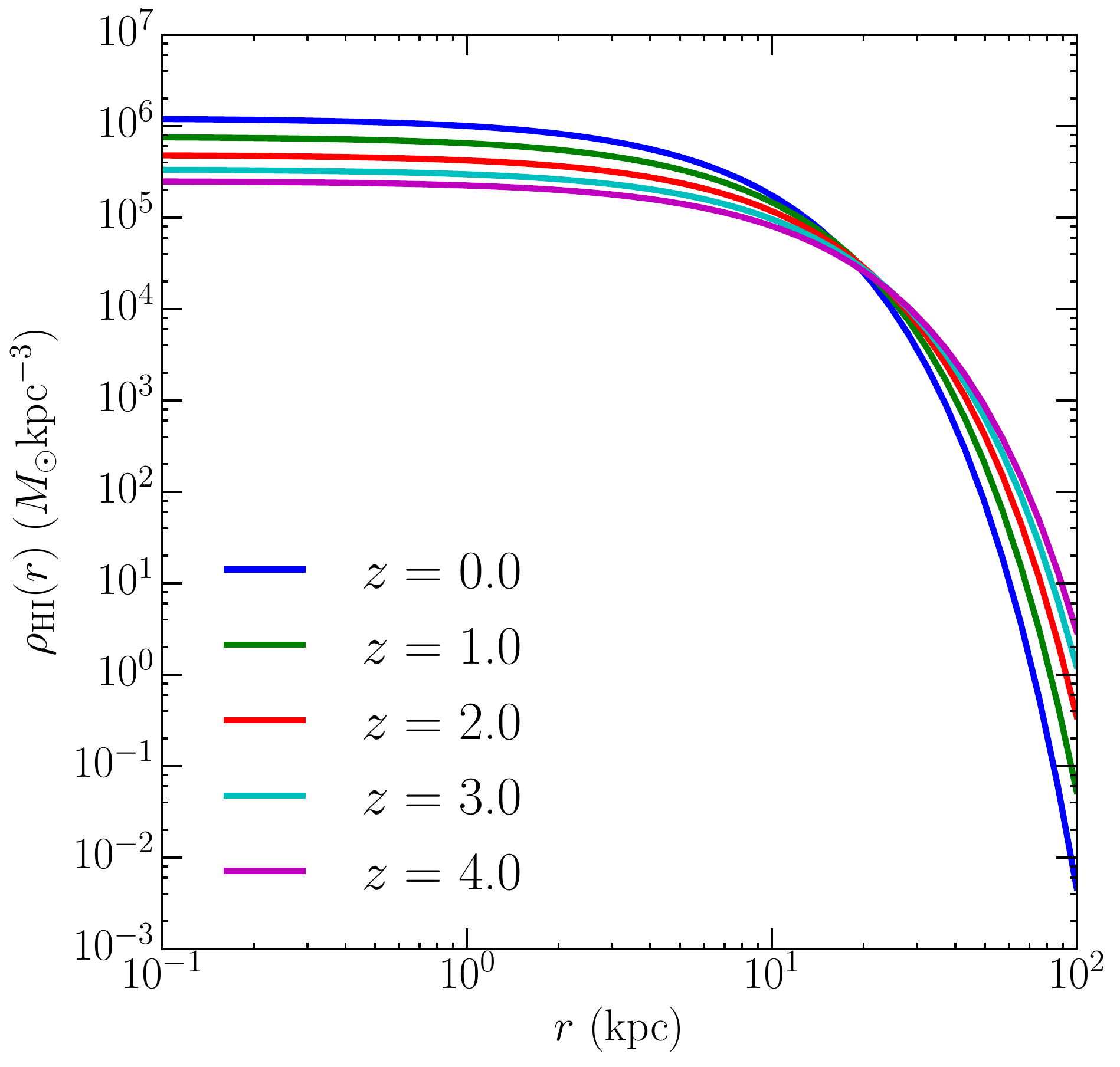}
\end{center}
\caption{The predicted evolution of the HI density profile $\rho_{\rm HI}(r)$ across redshifts, from the best-fit halo model. The figure shows the $\rho_{\rm HI} (r)$ for a halo of virial mass $M = 10^{12} h^{-1} M_{\odot}$ at the different redshifts.}
\label{fig:profilerelation}
\end{figure}
\item The model predictions are also in good agreement with the general trends in the covering fraction and impact parameter-column density relations of high-redshift DLA absorbers, as well as the surface mass densities observed in low-redshift exponential disks.

\item Although the model has some difficulty fitting the observed bias measurement at $z \sim 2.3$, the model predictions are consistent with results from imaging and other DLA surveys that favour DLAs at high-redshift to be hosted by faint dwarf galaxies \citep[e.g.,][]{cooke2015, fumagalli2014, fumagalli2015}, and the results of hydrodynamical simulations \citep[e.g.,][]{dave2013, rahmati2014} and cross-correlations \citep[e.g.,][]{bouche2005}. 

\item {The evolution in the present HIHM relation is passive and corresponds to the implicit redshift evolution of halo mass for a fixed cutoff virial velocity. From the approximate relation connecting the virial velocity to host halo mass $M$ at redshift $z$, $v_c(M,z) \propto M^{1/3} (1 + z)^{1/2}$, we see that at fixed virial velocity, the effective cutoff halo mass evolves roughly as $M_{\rm cutoff} \propto (1 + z)^{-3/2}$. This is shown in Fig. \ref{fig:massrelation} which indicates that the cutoff mass decreases with increasing redshift, and is also consistent with previous work \citep{hpar2016}}. 

\item We also note that evolving HIHM scenarios, i.e. modified functional forms of the $M_{\rm HI} - M$ relation with evolution in one, two, or all of the free parameters $\alpha$, $\beta$ and $v_{c,0}$, are not found to be statistically favoured over the base non-evolving scenario when the complete set of observations is taken into account  (although they may be favoured by particular subsets of the data). {Examples of this include fits to the $z \sim 2.3$ data only \citep[e.g.,][]{castorina2016, barnes2014} or using a subset of high-redshift ($z \sim 1 - 4$) DLA data alone, as done in an abundance matching analysis where we do not have access to the HI mass functions at redshifts $z > 0$ \citep[e.g.,][]{hpgk2016}}.

\end{enumerate}

In  future work, it will be useful to use the model results to build forecasts for upcoming 21-cm observations, to be measured by current and future experiments.  The inclusion of small-scale clustering enables the calculation of the $k$-dependence of the HI power spectra and its evolution as a function of redshift. The present framework is found to be in good agreement with the HI surface density profiles at low redshifts and DLA covering fractions and impact parameter observations at higher redshifts. It would be worthwhile to explore these relationships in greater detail in future analyses, by connecting the present model to stellar-halo mass relations. This would also be useful for comparing to the results of hydrodynamical simulations, especially for constraining the stellar-cold gas evolution with redshift.

\begin{table}
\centering
\caption{Summary of the best-fitting HI halo model and the free parameters.}
\label{table:final}
\begin{tabular}{|c|}
\hline
$M_{\rm HI} (M) = \alpha f_{H,c} M \left(M/10^{11} h^{-1} M_{\odot}\right)^{\beta} \exp\left[-\left(v_{c0}/v_c(M)\right)^3\right]$ \\
\\ $\rho_{\rm HI} (r) =\rho_0 \exp(-r/r_s)$;\\ 
\\ $c_{\rm HI}(M,z) \equiv R_v/r_s = c_{\rm HI, 0} \left(M/10^{11} M_{\odot} \right)^{-0.109} 4/(1+z)^{\gamma}$
 \\
\\
\hline 
\\
$c_{\rm HI,0} = 28.65 \pm 1.76$                                                                                                                                                                                                                                                                                                                                                                         \\
\\
$\alpha = 0.09 \pm 0.01$                                                                                                                                                                                                                                                                                                                                                                               \\
\\
log $v_{c,0} = 1.56 \pm 0.04$                                                                                                                                                                                                                                                                                                                                                                          \\
\\
$\beta = -0.58 \pm 0.06$                                                                                                                                                                                                                                                                                                                                                                               \\
\\
$\gamma = 1.45 \pm  0.04$                                                                                                                                                                                                                                                                                                                                                                              
\\
\hline
\end{tabular}
\end{table}

\section*{Acknowledgements}
{We thank the anonymous referee for a helpful report.} We thank Girish Kulkarni and Ali Rahmati for useful discussions, and Joel Akeret for help and support with the \textsc{cosmohammer} package. The research of HP is supported by the Tomalla Foundation. The simulations used in this work were carried out on the Monch cluster of the Swiss Supercomputing Center CSCS.

\appendix

\section{Modified NFW profile}
\label{sec:appendix}

\begin{figure*}
\begin{center}
\includegraphics[scale=0.6, width = \textwidth]{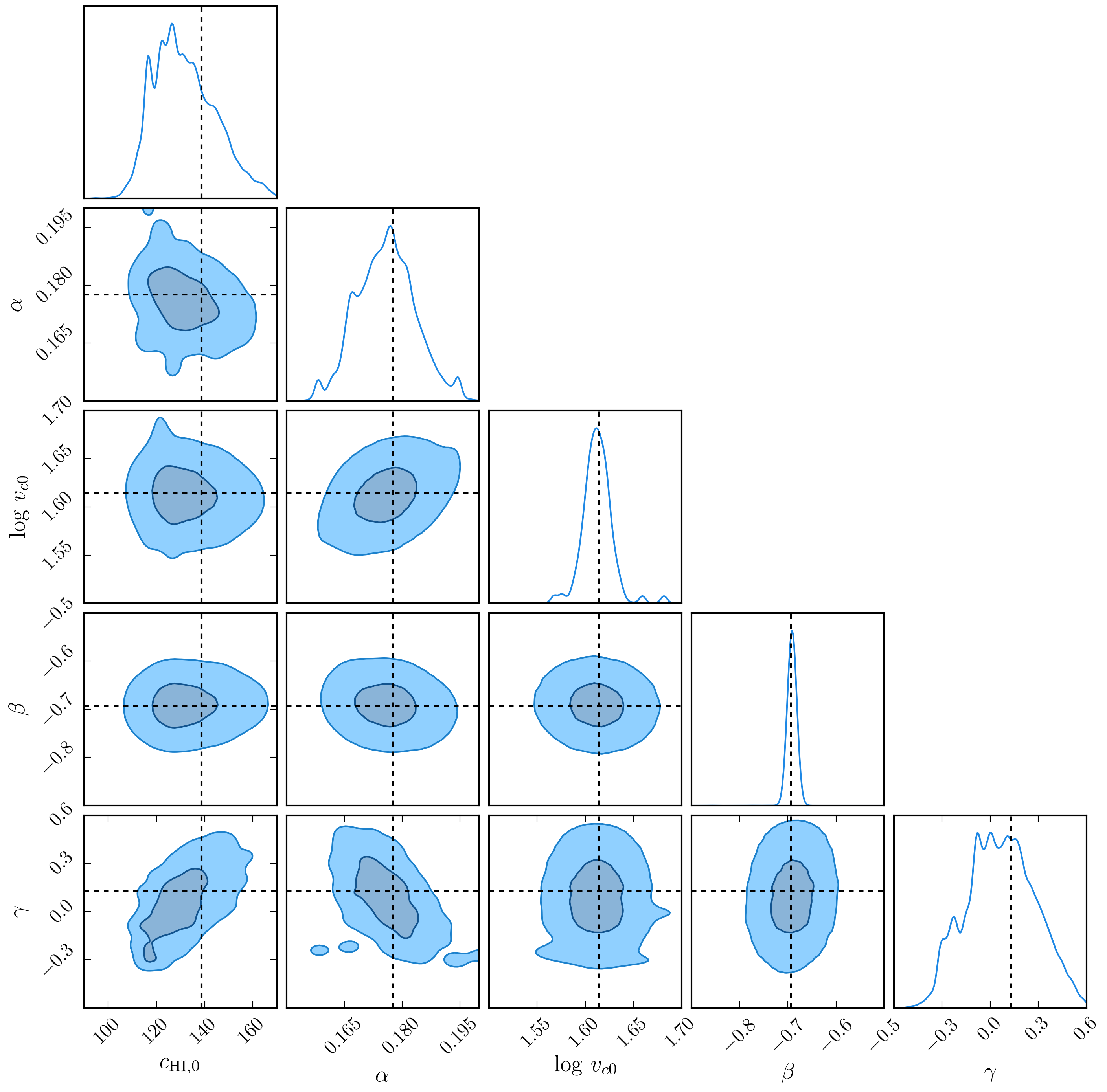}
\end{center}
\caption{Parameter space of the MCMC analysis with an altered NFW profile. Contours indicate 1- and 2$\sigma$ confidence levels. The crosshairs indicate the maximum likelihood estimate of the parameters (the best-fitting value). The marginal distributions for the parameters are shown in the diagonal panels.}
\label{fig:cornerplotnfw}
\end{figure*}

In this Appendix, we consider the modifications to the analysis when an altered NFW profile (e.g. \citet[][Paper I]{barnes2014} is used,  instead of the exponential profile considered in the main text.  This form is found to be a good fit to multiphase halo gas in simulations at high redshifts \citep{maller2004}:
\begin{equation}
\rho_{\rm HI} (r) = \frac{\rho_0 r_s^3}{(r + 0.75 r_s) (r+r_s)^2}
\label{rhodefnfw}
\end{equation}
The form of the HI- halo mass relation remains the same as in the main text.
For quantifying clustering, we used the normalized Fourier transform of the profile, given by:
\begin{equation}
u(k|M) = \frac{4 \pi \rho_0 r_s^3 u_1(k|M)}{M_{\rm HI} (M)}
\end{equation}
where
\begin{eqnarray}
u_1(k|M) &=& -(-12 \mathrm{Ci}(0.75  k r_s) \sin(0.75  k r_s) \nonumber \\
&+& 12 \mathrm{Ci}((0.75 + c_{\rm HI})  k r_s) \sin(0.75  k r_s) \nonumber \\
&+& \mathrm{Ci}((1 + c_{\rm HI})  k r_s) (4  k r_s \cos( k r_s) - 12 \sin( k r_s))\nonumber \\
 &+& \mathrm{Ci}( k r_s) (-4  k r_s \cos( k r_s) + 12 \sin( k r_s)) \nonumber \\
 &-& 4 \sin(c_{\rm HI}  k r_s)/(1. +c_{\rm HI})\nonumber \\
  &+& 12 \cos(0.75  k r_s) \mathrm{Si}(0.75  k r_s) - 12 \cos( k r_s) \mathrm{Si}( k r_s) \nonumber \\
  &-& 4  k r_s \sin( k r_s) \mathrm{Si}( k r_s) \nonumber \\
  &-& 12 \cos(0.75  k r_s) \mathrm{Si}((0.75 + c_{\rm HI})  k r_s) \nonumber \\
  &+& \mathrm{Si}((1 + c_{\rm HI})  k r_s) (12 \cos( k r_s) \nonumber \\
  &+& 4  k r_s \sin( k r_s)))/ (k r_s) 
\end{eqnarray}
with
\begin{equation}
\mathrm{Ci} (x) = - \int_x^{\infty} \frac{\cos t}{t} ; \ \  \mathrm{Si} (x) =  \int_0^{x} \frac{\sin t}{t}
\end{equation}
Repeating the MCMC analysis in the main text for this form of the profile with the same set of data, we find the parameter space of the results as shown in Fig. \ref{fig:cornerplotnfw}. As in the main text, the best-fitting values and their $1\sigma$ errors are summarized in Table \ref{table:bestfitnfw}.

\begin{figure}
\begin{center}
\includegraphics[scale=1, width = \columnwidth]{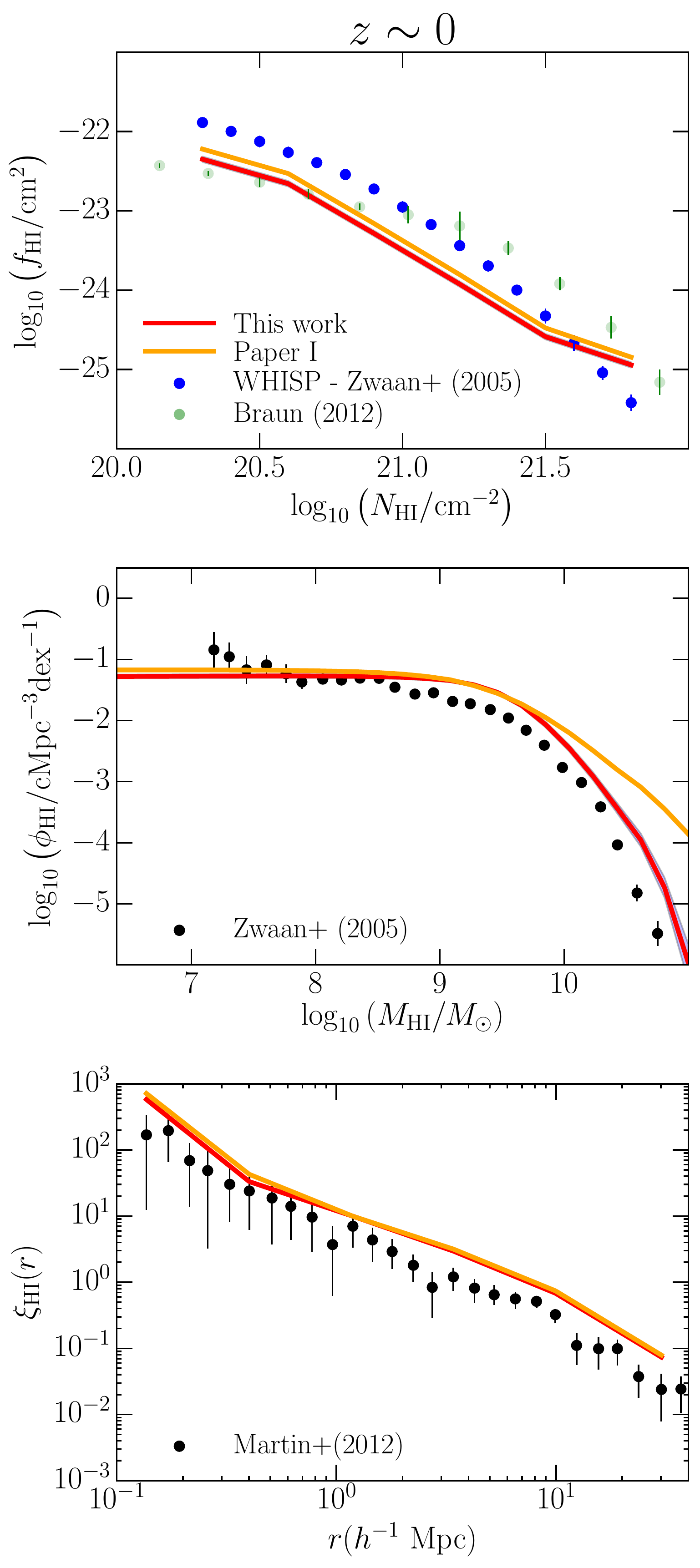}
\end{center}
\caption{Data and model predictions at $z \sim 0$, with the modified NFW profile. From top to bottom: (a) the column density distribution $f_{\rm HI}$ as measured from the WHISP \citep{zwaan2005a} and the \citet{braun2012} surveys, (b) the HI mass function $\phi_{\rm HI}$ from the results of the HIPASS \citep{zwaan05}  survey, and (c) the correlation function of HI-selected galaxies, $\xi_{\rm HI}(r)$ from the ALFALFA survey \citep{martin12}. The results from using the Paper I best-fit parameters are indicated in orange.}
\label{fig:clustermfnfw}
\end{figure}

\begin{figure*}
\begin{center}
\includegraphics[scale=1, width = \textwidth]{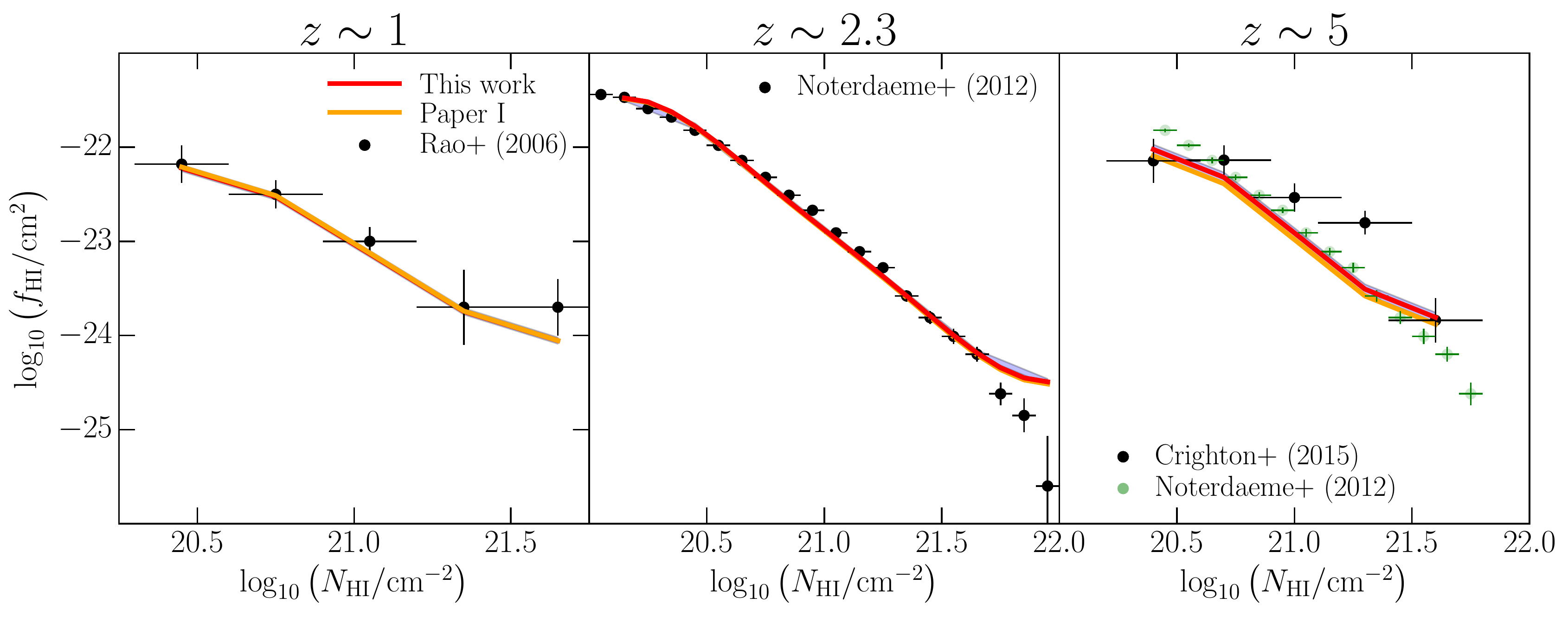}
\end{center}
\caption{The DLA column density distribution at redshifts $z > 0$, compared to the model predictions with the modified NFW profile. From left to right: the column density distribution for DLAS at $z \sim 1$, $z \sim 2.3$, and $z \sim 5$, compared with with the observations of \citet{rao06, noterdaeme12, crighton2015} respectively. The results on using the Paper I best-fit parameters are indicated in orange.}
\label{fig:columndensitynfw}
\end{figure*}

\begin{figure}
\begin{center}
\includegraphics[scale=1, width = \columnwidth]{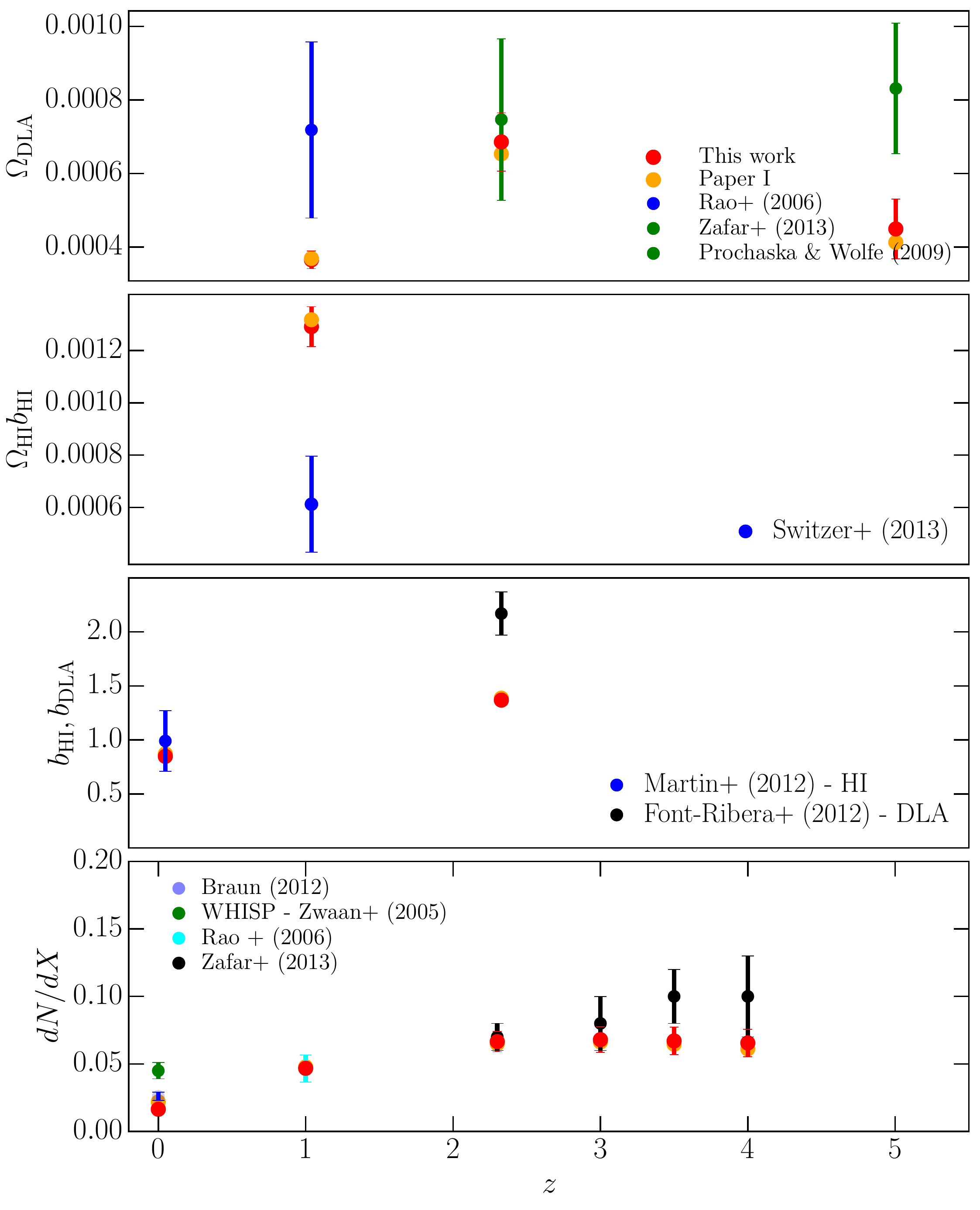}
\end{center}
\caption{From top to bottom: The observables $\Omega_{\rm DLA}$,  $\Omega_{\rm HI} b_{\rm HI}$ from intensity mapping experiments, the bias measurements $b_{\rm HI}$ and $b_{\rm DLA}$, and the DLA incidence $dN/dX$. Also shown are the model predictions (red) with the modified NFW profile, at the corresponding redshifts. The results using the Paper I best-fit parameters are indicated in orange.}
\label{fig:omegabiasdndxnfw}
\end{figure}

\begin{figure}
\begin{center}
\includegraphics[scale=0.5, width = \columnwidth]{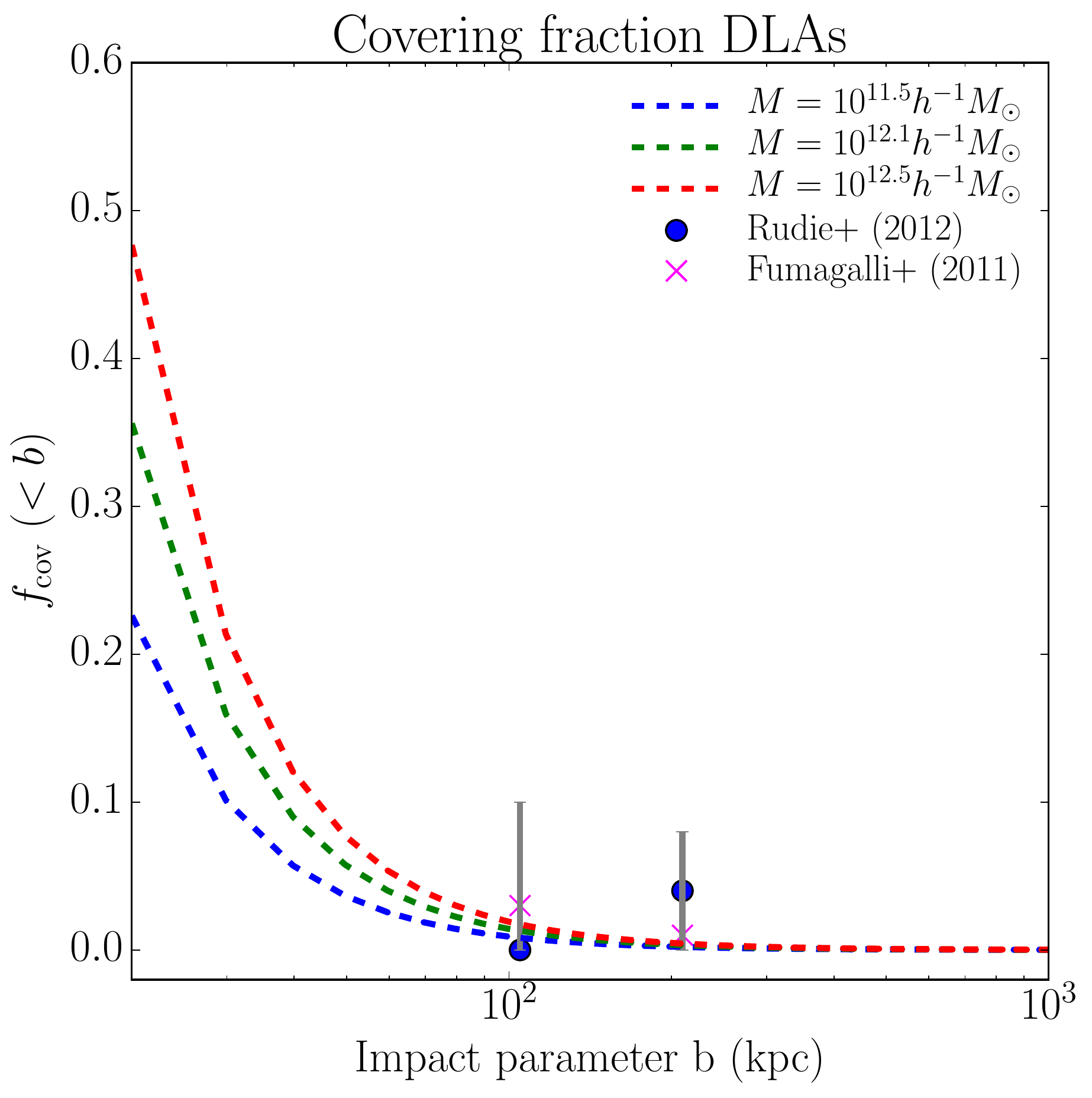}
\end{center}
\caption{The covering fraction as a function of impact parameter, as predicted by the HI halo model with a modified NFW profile at $z \sim 2.5$, along with the observations from the Keck Baryonic Spectroscopic Survey \citep[KBSS;][blue circles]{rudie2012} and the cosmological zoom-in simulations of \citet[][pink crosses]{fumagalli2011}.}
\label{fig:covfractionnfw}
\end{figure}

\begin{table}
\centering
\caption{Priors and best-fitting values of the free parameters in the HI halo model with an altered NFW profile.}
\label{table:bestfitnfw}
\begin{tabular}{llll}
\hline 
Parameter  & Prior & Best-fit \& Error (1$\sigma$) \\
\hline \\
$c_{\rm HI}$ &    [20, 400]   &   $139 \pm 13$               \\
$\alpha$     & [0.05, 0.5]      &    $ 0.176 \pm 0.007$                 \\
log $v_{c,0}$    & [1.30, 1.90]      &  $1.61 \pm 0.02$                \\
$\beta$      &   [-1,3]    &    $-0.69 \pm 0.03$               \\
$\gamma$     &  [-0.9,2]     &       $0.13 \pm 0.20$       \\
\hline \\    
\end{tabular}
\end{table}
The model predictions with this form of the profile are compared to the observations in Figs. \ref{fig:clustermfnfw}, \ref{fig:columndensitynfw},  and   \ref{fig:omegabiasdndxnfw} respectively. 
We find that the model reproduces the high-redshift data reasonably well but leads to a tension between the column density distribution and the HI mass function at $z \sim 0$ \citep[also noticed in previous work, Paper I and][]{hpgk2016}. This indicates that the exponential profile may better describe the column density and clustering observations at $z \sim 0$,  as compared to the altered NFW profile. The predictions from the best-fitting parameters of Paper I are indicated on each figure in orange, showing that the results from Paper I are an overall good fit to the extended data sample. At high redshifts, the model predictions also reproduce the trends in the impact parameter-covering fraction of DLAs seen in observations (Fig. \ref{fig:covfractionnfw}) and the  impact parameter-column density relations \citep[e.g.,][]{rao2011,krogager2012, peroux2013}.\footnote{We also note that the model predictions somewhat overestimate the value of $\Omega_{\rm HI} b_{\rm HI}$ at $z \sim 1$. However, recent work suggests that using improved techniques of galactic foreground removal  may lead to an increase in the observed value of $\Omega_{\rm HI} b_{\rm HI}$ by up to an order of magnitude \citep{wolz2015}.}

\bibliographystyle{mnras} 
\def\aj{AJ}                   
\def\araa{ARA\&A}             
\def\apj{ApJ}                 
\def\apjl{ApJ}                
\def\apjs{ApJS}               
\def\ao{Appl.Optics}          
\def\apss{Ap\&SS}             
\def\aap{A\&A}                
\def\aapr{A\&A~Rev.}          
\def\aaps{A\&AS}              
\def\azh{AZh}                 
\def\baas{BAAS}
\def\jcap{JCAP}
\def\jrasc{JRASC}             
\def\memras{MmRAS}
\def\na{New Astronomy}
\def\nat{Nature}
\def\mnras{MNRAS}             
\def\pra{Phys.Rev.A}          
\def\prb{Phys.Rev.B}          
\def\prc{Phys.Rev.C}          
\def\prd{Phys.Rev.D}          
\def\prl{Phys.Rev.Lett}       
\def\pasp{PASP}               
\def\pasj{PASJ}
\def\physrep{Phys. Repts.}
\def\qjras{QJRAS}             
\def\skytel{S\&T}             
\def\solphys{Solar~Phys.}     
\def\sovast{Soviet~Ast.}      
\def\ssr{Space~Sci.Rev.}      
\def\zap{ZAp}                 
\let\astap=\aap
\let\apjlett=\apjl
\let\apjsupp=\apjs

\bibliography{mybib}

\end{document}